\documentclass[openright, a4paper, 10pt]{article}
\usepackage[utf8]{inputenc}
\usepackage[T1]{fontenc}
\usepackage[USenglish, german]{babel}
\usepackage{mathtools}
\usepackage{amssymb}
\usepackage{tikz,pgfplots}
\usetikzlibrary{chains}
\usepgfplotslibrary{colormaps}
\pgfplotsset{
    colormap={slategraywhite}{
        rgb255=(109,5,41)
        rgb255=(163,80,44)
        rgb255=(169,160,98)
    },
}
\pgfplotsset{
    colormap={unityjet}{
        rgb255=(107,2,40)
        rgb255=(163,80,44)
        rgb255=(169,160,98)
        rgb255=(132,155,98)
        rgb255=(70,126,102)
        rgb255=(62,54,99)                
    },
}

\pgfplotsset{
    colormap={baltim}{
        rgb255=(11,55,134)
        rgb255=(11,122,176)
        rgb255=(160,184,101)
        rgb255=(242,181,56)
        rgb255=(235,116,49)
        rgb255=(217,31,30)                
    },
}

\pgfplotscreateplotcyclelist{mycolorlist}{%
blue,every mark/.append style={fill=blue!80!black},mark=*\\%
red,every mark/.append style={fill=red!80!black},mark=square*\\%
brown!60!black,every mark/.append style={fill=brown!80!black},mark=otimes*\\%
black,mark=star\\%
blue,every mark/.append style={fill=blue!80!black},mark=diamond*\\%
red,densely dashed,every mark/.append style={solid,fill=red!80!black},mark=*\\%
brown!60!black,densely dashed,every mark/.append style={
solid,fill=brown!80!black},mark=square*\\%
black,densely dashed,every mark/.append style={solid,fill=gray},mark=otimes*\\%
blue,densely dashed,mark=star,every mark/.append style=solid\\%
red,densely dashed,every mark/.append style={solid,fill=red!80!black},mark=diamond*\\%
}

\usetikzlibrary{patterns}
\usetikzlibrary{3d}
\usetikzlibrary{shapes,arrows,positioning,calc}
\usetikzlibrary{calc}
 \usetikzlibrary{matrix}

\usepackage{listofitems,readarray}
\tikzset{
  pics/carc/.style args={#1:#2:#3}{
    code={
      \draw[pic actions,fill=gray] (#1:#3) arc(#1:#2:#3);
    }
  }
}
\pgfmathdeclarefunction{lg10}{1}{ \pgfmathparse{ln(#1)/ln(10)}}
 \usepgfplotslibrary{colormaps}
\pgfplotsset{
    colormap={slategraywhite}{
        rgb255=(112,128,144)
        rgb255=(255,159,101)
    },
}
\tikzstyle{decision} = [diamond, draw, fill=blue!20, 
    text width=10.em, text badly centered, node distance=1cm, inner sep=0pt]
\tikzstyle{block} = [rectangle, draw, fill=blue!20, 
    text width=25em, text centered, rounded corners, minimum height=1em]
\tikzstyle{line} = [draw, -latex']
\tikzstyle{cloud} = [draw, ellipse,fill=red!20, node distance=3cm,
    minimum height=1.5em]

\definecolor{mylightred}{RGB}{211,79,73}
\definecolor{mydarkred}{RGB}{199,44,38}
\definecolor{mylightgreen}{RGB}{78,153,67}
\definecolor{mydarkgreen}{RGB}{43,129,33}
\definecolor{mylightpurple}{RGB}{150,107,178}
\definecolor{mydarkpurple}{RGB}{126,78,160}
\definecolor{mylightblue}{RGB}{49,101,205}
\definecolor{mydarkblue}{RGB}{20,92,205}

\tikzset{
  juliadot/.style args={#1,#2}{shape=circle,line width=0.03ex,minimum width=0.4ex,fill=#1,draw=#2}
}
\newcommand\julialetter[1]{{\strut\fontfamily{cmss}\bfseries\selectfont{#1}}}
\DeclareRobustCommand\julia{%
\begin{tikzpicture}[baseline=0mm, every node/.style={inner sep=0mm, outer sep=0mm}]
\node[anchor=base]        (j) at (0,0) {\julialetter{\j}};
\node[anchor=base, right=0ex of j] (u) {\julialetter{u}};
\node[anchor=base, right=0ex of u] (l) {\julialetter{l}};
\node[anchor=base, right=0ex of l] (i) {\julialetter{\i}};
\node[anchor=base, right=0ex of i] (a) {\julialetter{a}};
\path let \p1 = (j) in node[juliadot={mylightblue,mydarkblue}] (bluedot) at (\x1+0.02ex,1.4ex) {};
\path let \p1 = (i) in node[juliadot={mylightred,mydarkred}] (reddot) at (\x1,1.4ex) {};
\path let \p1 = (reddot) in node[juliadot={mylightpurple,mydarkpurple}] (purpledot) at (\x1+0.5ex,\y1) {};
\path let \p1 = (reddot) in node[juliadot={mylightgreen,mydarkgreen}] (greendot) at (\x1+0.25ex,\y1+0.42ex) {};
\end{tikzpicture}%
}
\usepackage[noprefix]{nomencl}

\renewcommand{\nomgroup}[1]{\medskip}
\makenomenclature

\usepackage{booktabs}

\definecolor{shadedcolor}{gray}{0.8}
\definecolor{TFFrameColor}{gray}{.8} % frame color
\definecolor{TFTitleColor}{gray}{0} % title font color
\usepackage{framed}

\usepackage{hyperref}
\hypersetup{
    colorlinks,
    citecolor=black,
    filecolor=black,
    linkcolor=black,
    urlcolor=black
}
\usepackage{tabularx}
\usepackage{parskip}

\usepackage{fancyhdr}
\usepackage[toctitles]{titlesec}
\titleformat{\chapter}[hang]{\bfseries\huge}{\thechapter}{2pc}{}
\titlelabel{\thetitle\quad}   % For consistency in all headings

\makeatother
\let\sectionOld\section
\renewcommand\section[2][\empty]{%
	\boldmath\sectionOld[#1]{#2}\unboldmath%
}

\usepackage{enumerate}

\usepackage{siunitx}
\usepackage{units}
\usepackage{upgreek}
\usepackage{tensor}
\usepackage{listings}

\newcommand{\eq}[1]{\begin{align}#1\end{align}}

\newcommand{\mbf}{\mathbf}
\newcommand{\bs}{\boldsymbol}

\newcommand{\vect}{\pmb} %Vector (boldface)
\newcommand{\vcomp}[1]{#1} %Vector component op
\newcommand{\tens}{\mbf} % (usually 2nd order) Tensor
\newcommand{\tcomp}{\text} % Tensor component op
\newcommand{\II}{\mbf{I}} %identity tensor order 2


\usepackage[hang]{footmisc}

\let\originalleft\left
\let\originalright\right
\renewcommand{\left}{\mathopen{}\mathclose\bgroup\originalleft}
\renewcommand{\right}{\aftergroup\egroup\originalright}

\begin{document}

	\selectlanguage{USenglish}
	% if work should have blank page first (in a bound book for example)
	%\thispagestyle{empty}
	%\mbox{}
	%\newpage
	
%	% write out a custom title page
%	\begin{titlepage}
%	\thispagestyle{fancy}
%		\fancyhead{}
%		\fancyfoot{}
%		\fancyhead[L]{\hspace{1cm}\includegraphics[scale=.1]{luh}}
%		\fancyhead[R]{\includegraphics[scale=.12]{ibnm}\hspace{1cm}}
%		\renewcommand{\headrulewidth}{0pt}
%		\headheight 56pt
%		\onehalfspacing
%		\begin{center}
%			\hbox{}
%			\vspace{1.5cm}
%			
%			{\large \bf Course Project}\\
%			\vspace{1cm}
%			
%			{\Large \sc Computational Contact Mechanics}
%			\par
%			\vspace{1.5cm}
%			{\large \emph{Robert Lee Gates}}\\
%			{\scriptsize Matr.-Nr.: 2708610}
%			\par
%			\vfill
%			
%			{\sc Gottfried Wilhelm Leibniz University Hannover}
%			\par
%			\vspace{.5cm}
%			Faculty of Civil Engineering and Geodetic Science
%			\par
%			\vspace{.5cm}
%			Institute of Mechanics and Computational Mechanics
%			\par
%			\vspace{1cm}
%			\emph{Univ.-Prof. Dr.-Ing. Udo Nackenhorst}\\
%			\emph{Dr. Wenzhe Shan}
%			\par
%			\vspace{1.5cm}
%			\monthname\ \number\year
%			\vspace{.7cm}
%		\end{center}
%		
%		\newpage
%		%\thispagestyle{empty}
%		%\mbox{}
%		%\newpage
%	\end{titlepage}

%	\singlespacing
	% reset standard fancyhdr
	% remove "chapter" from title, remove trailing dot after section number
	%\renewcommand{\chaptermark}[1]{\markboth{\MakeUppercase{#1}}{}}
	\pagestyle{fancy}
	
	\lhead[\MakeUppercase{IBNM Preprint 09/2015}]{\MakeUppercase{IBNM Preprint 09/2015}}
	\rhead[]{}
	%\lfoot[\thepage]{}
	%\rfoot[]{\thepage}

%	\newpage
%	\thispagestyle{empty}
%	\mbox{}
%	\singlespacing
%	\begin{bottompar}
%		{\footnotesize Copyright (2012) Robert Lee Gates. \\
%		The author reserves all rights to this work and the computer code contained herein. Redistribution of any kind requires the express written consent of the author, except where explicitly noted otherwise.}
%	\end{bottompar}
%	\onehalfspacing
	% front matter
	\pagenumbering{roman}

	% set the table of contents to number all up to subsubsections. see: http://en.wikibooks.org/wiki/LaTeX/Document_Structure
	\setcounter{tocdepth}{3}

	% abstract
	%\selectlanguage{USenglish}

%	\begin{abstract}
%	\par
%	\parindent=1cm
%	\cite{Nakamachi2007}
%	
%	
%	
%	\end{abstract}
	
	\parindent=.5cm
	\selectlanguage{USenglish}

	\selectlanguage{USenglish}
	
	\parindent=.5cm
	%\thispagestyle{empty}
	%\hbox{}
	%\newpage{}
	\pagestyle{plain}
	% make a toc
	
%	\tableofcontents
	
	% list of figures, list of tables, nomenclature
%	\cleardoublepage
%	\phantomsection
%	\addcontentsline{toc}{chapter}{List of Figures}
%	\listoffigures
%	
%	\cleardoublepage
%	\phantomsection
%	\addcontentsline{toc}{chapter}{List of Tables}
%	\listoftables
	
%	\cleardoublepage
%	\phantomsection
%	\addcontentsline{toc}{chapter}{Notation}
%	\include{chapters/nomenclature}
%	\printnomenclature[2cm]
	
%	\newpage
%	\thispagestyle{empty}
%	\mbox{}
	% begin main matter, include chapters
	\cleardoublepage
	\pagenumbering{arabic}	
	\pagestyle{fancy}

	\thispagestyle{plain}

\begin{center}
	{\bf IBNM Preprint 05/2020}\\
	\vspace{.5cm}
	{\LARGE Technical Report: Virtual X-ray imaging for higher-order finite element results}\\
	\vspace{.8cm}
	{Maximilian Bittens\footnote{maximilian.bittens@ibnm.uni-hannover.de, Institute of Mechanics and Computational Mechanics (IBNM), Gottfried Willhelm Leibniz Universität Hannover, Appelstr. 9A, D-30167 Hannover}}\\
	\vspace{.2cm}
\end{center}
\vspace{1cm}

\noindent
\begin{center}{\bf Abstract}\medskip\end{center}
This work describes and demonstrates the operation of a virtual X-ray algorithm operating on finite-element post-processing results which allows for higher polynomial orders in geometry representation as well as density distribution. A nested hierarchy of oriented bounding boxes is used for preselecting candidate elements undergoing a ray-casting procedure. The exact intersection points of the ray with the finite element are not computed, instead the ray is discretized by a sequence of points. The element-local coordinates of each discretized point are determined using a local Newton iteration and the resulting densities are accumulated. This procedure results in highly accurate virtual X-ray images of finite element models.
\vspace{.5cm}
\section{Introduction}
\label{ch:xraysimu}
Finite element analyses in orthopaedic biomechanics provide deep insights into the physics of the human body. For such analyses to represent a meaningful complement to purely medical research, it can be useful to transfer the results into a medical imaging format, such as a radiographic or tomographic image. For example, in computational bone remodeling there are two major advantages to this approach: (1) the internal bone mineral density distribution is hard to infer from finite element post-processing results, since only the boundary of the specimen is immediately visible and (2) also people without technical background in finite elements, e.g. medicines, can evaluate the results in that way.\\
The simulation of X-ray images from a polygonal mesh is a well-researched topic and there are a number of simulation codes available, such as described in \cite{freud2006fast}, \cite{baro1995penelope} and \cite{sujar2017gvirtualxray}, for example. \cite{baro1995penelope} use Monte Carlo simulation to generate realistic images, while \cite{freud2006fast} and \cite{sujar2017gvirtualxray} use fast raycasting algorithms for that purpose. Whilst \cite{sujar2017gvirtualxray} account for density distribution with higher polynomial order, to the authors knowledge, there is no work available which operates directly onto finite element results and accounts for higher polynomial orders in both density distribution and geometry representation. For this purpose a novel approach will be presented in this approach.%, outlined as follows:
%\section{X-ray setup}
\section{Method description}
In order to obtain an x-ray image of an arbitrary finite element result, the approved method of sending rays through the finite element mesh and subsequently integrating the quantity of interest along these rays will be adopted, here. Since an intersection test of each ray with every finite element would be computationally expensive, as a first step and for a given ray, the amount of finite element candidates for the intersection test have to be reduced. For this purpose, a tree-structure of nested bounding-boxes is employed, whereby the elements contained in a bounding box are bisected in the next hierarchical level. Since intersection tests with bounding boxes are very cheap, the tree of bounding boxes can be efficiently used to reduce the amount of candidate elements. Since, even with this method, the determination of the exact points of intersection of the ray and the finite element will be computationally expensive if performed for a large amount of rays or a very large finite element mesh, numerical integration is introduced at this point. Thus, it is only necessary to test whether a discrete point is inside a finite element. As soon as the positional relationship of the point and the finite element is known in terms of element coordinates, it is simple to obtain the quantity of interest at the discrete point with the help of shape functions from a finite element postprocessing result.

\section{Hierarchical-structured oriented bounding boxes}
\label{bittens_sec:hsobb}
%%%%%%%%%%%%%%%%%%%%%%%%%%%%%%%%%%%%%%%%%%%%%%%%
%%%%%%%%%%%%%%%%%%%%%%%%%%%%%%%%%%%%%%%%%%%%%%%%
\begin{figure}[htb]
\begin{center}
    \begin{minipage}[t]{.275\textwidth}
        \centering
        \includegraphics[width=\textwidth]{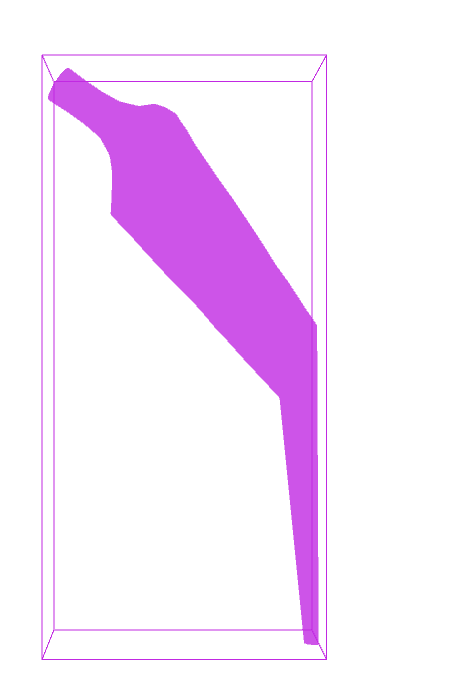}
    \end{minipage}
    \hspace{.5cm}
    \begin{minipage}[t]{.275\textwidth}
        \centering
        \includegraphics[width=\textwidth]{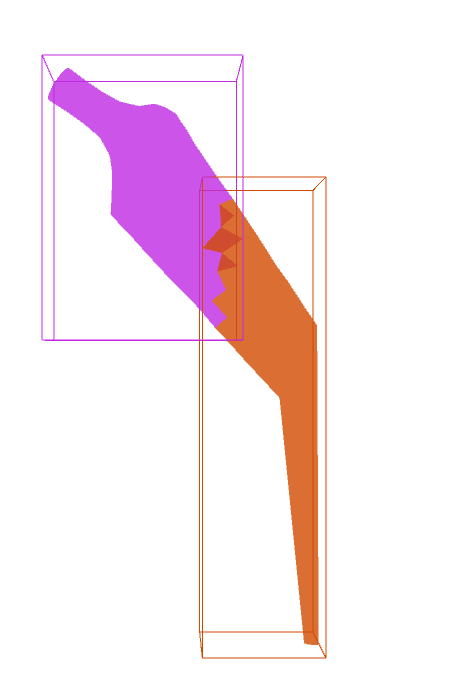}
    \end{minipage}
        \hspace{.5cm}
    \begin{minipage}[t]{.275\textwidth}
        \centering
        \includegraphics[width=\textwidth]{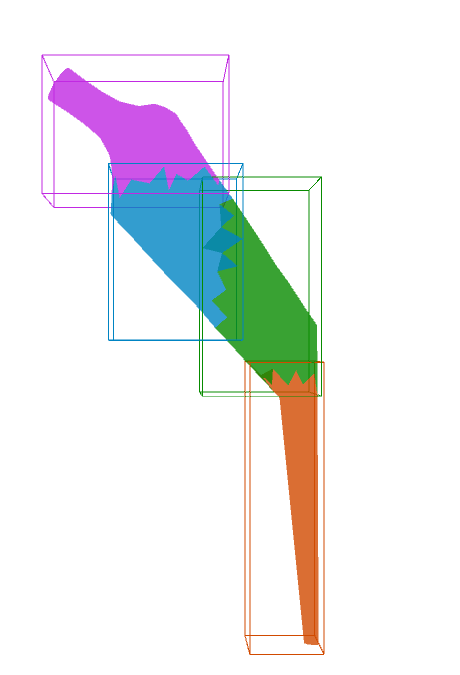}
    \end{minipage}\\
     \begin{minipage}[t]{.275\textwidth}
        \centering
        \includegraphics[width=\textwidth]{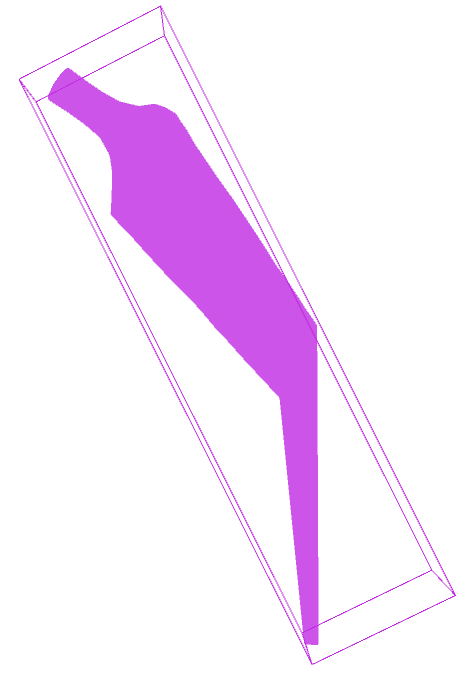}
    \end{minipage}
        \hspace{.5cm}
    \begin{minipage}[t]{.275\textwidth}
        \centering
        \includegraphics[width=\textwidth]{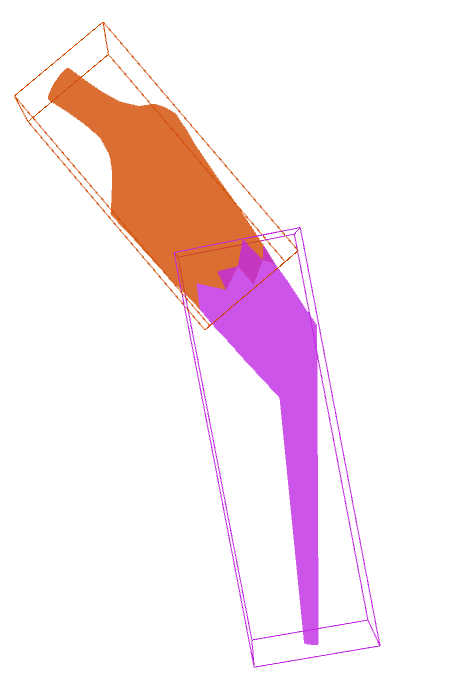}
    \end{minipage}
        \hspace{.5cm}
    \begin{minipage}[t]{.275\textwidth}
        \centering
        \includegraphics[width=\textwidth]{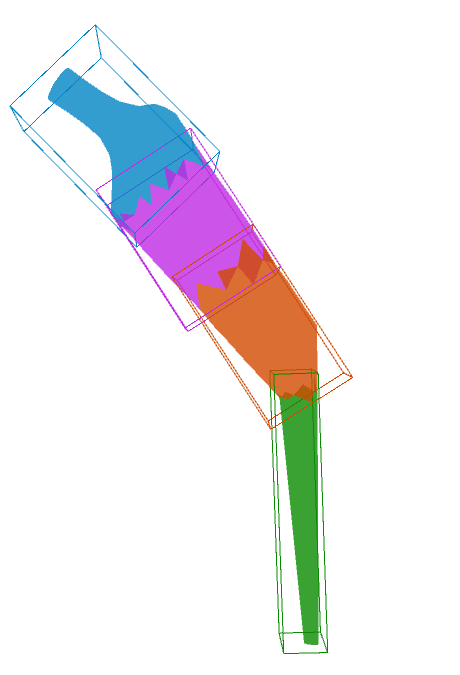}
    \end{minipage} 
    \end{center}
    \caption{AABBs (upper pictures) vs. OBBs (lower pictures).}
    \label{bittens_fig:AABBvsOBB}
\end{figure}\noindent
A bounding box for a collection of objects, such as an axis aligned bounding box (AABB) or an oriented bounding box (OBB)  is a closed volume, that completely contains the collection of objects. In \cite{gottschalk1996obbtree} a tree-structure of hierarchical OBB's is introduced. Therein, principle component analysis \cite{jolliffe2011principal} is used to create a set of nested tight-fitting bounding boxes. This method could also be used efficiently for contact detection in finite element method \cite{gottschalk1996obbtree}.
%%%%%%%%%%%%%%%%%%%%%%%%%%%%%%%%%%%%%%%%%%%%%%%%
\subsection{Principle component analysis}
\label{bittens_sec:pca}
%%%%%%%%%%%%%%%%%%%%%%%%%%%%%%%%%%%%%%%%%%%%%%%%
Let $k=1,...,n$ and $s_k$ be a linear triangle or so-called 2-simplex with vertices ${\hat{\vect{p}}}^k$, ${\hat{\vect{q}}}^k$ and ${\hat{\vect{r}}}^k$ in $\mathbb{R}^3$. The centroid and the area of each simplex $s_k$ in $\mathbb{R}^3$ may be calculated by
\begin{align}
\vect{c}^k &= \frac{1}{3}\left( \hat{\vect{p}}^k + \hat{\vect{q}}^k + \hat{\vect{r}}^k \right)\quad\text{and}\\
A^k&=\sqrt{\gamma^k\left(\gamma^k-||\hat{\vect{p}}^k-\hat{\vect{q}}^k||\right)\left( \gamma^k - ||\hat{\vect{q}}^k - \hat{\vect{r}}^k || \right)\left( \gamma^k - || \hat{\vect{r}}^k-\hat{\vect{p}}^k || \right)}\,,
\end{align}
subsequently, with 
\begin{align}
\gamma^k=\frac{1}{2}\left(||\hat{\vect{p}}^k-\hat{\vect{q}}^k||+||\hat{\vect{q}}^k - \hat{\vect{r}}^k ||+|| \hat{\vect{r}}^k-\hat{\vect{p}}^k ||\right)\,.
\end{align}
The (area-)weighted center of all simplices $s_k$ can then be given by
\begin{align}
\boldsymbol{\mu} =  \frac{\sum\limits_k A^k \vect{c}^k}{\sum\limits_k A^k} \,,
\label{eq:areaweightedcenter}
\end{align}
which is independent of the distribution of the vertices in space. Using zero-centered centroids
\begin{align}
\bar{\vect{c}}^k = \sqrt{A^k}\left(\vect{c}^k - \boldsymbol{\mu}\right)\,,
\end{align}
the definition of the covariance matrix is given by
\begin{align}
\mathcal{C}_{ij} = \frac{1}{n-1} \sum\limits_k\bar{\vcomp{c}}^k_i \bar{\vcomp{c}}^k_j\,,
\end{align}
which is a symmetric positive (semi-)definite $3\times3$-matrix and thus owns real eigenvalues and, due to spectral theorem, an orthonormal basis $\mathcal{B}'$  in $\mathbb{R}^3$ consisting of the three eigenvectors out of which two point in the direction of maximum and minimum variance \cite{jolliffe2011principal}. A change of basis $\mathcal{T}^\mathcal{B}_{\mathcal{B}'}$ can be performed by
\begin{align}
\mathcal{T}^\mathcal{B}_{\mathcal{B}'} = (\mathcal{B}')^{-1}  \mathcal{B} = (\mathcal{B}')^{-1}  \II =(\mathcal{B}')^{-1} = \mathcal{T}_{\mathcal{B}'} \,,
\end{align}
with identity matrix $\II$, since $\mathcal{B}$ is the standard basis for a Euclidean space. This procedure is easily extendable to higher dimensions or different primitives by specifying suitable definitions for $\mathbf{c}^k$ and $A^k$ for the $n$-simplex with $n>2$ or e.g.~the $n$-cube.
%%%%%%%%%%%%%%%%%%%%%%%%%%%%%%%%%%%%%%%%%%%%%%%%
\subsection{Oriented bounding boxes}
%%%%%%%%%%%%%%%%%%%%%%%%%%%%%%%%%%%%%%%%%%%%%%%%
Let $S=\left\{ s_k \right\}_{k=1,..,n}$ be a set of 2-simplices or triangles and 
\begin{align}
V = \bigcup\limits_{s_k\in S} \left\{  \hat{\vect{p}}^k,\, \hat{\vect{q}}^k,\, \hat{\vect{r}}^k \right\}
\end{align}
the set of 0-simplices or vertices contained in $S$. A convenient way to define a bounding box as an axis-aligned rectangular cuboid is given by the tuple
\begin{align}
B(V) &= \left( P^{\text{min}}(V),\, P^{\text{max}}(V) \right)\,, \quad \text{with} \\
P^{\text{min}}(V) &= (\text{min}_{x_1}(V),\text{min}_{x_2}(V),\text{min}_{x_3}(V)) \quad \text{and} \\
P^{\text{max}}(V) &= (\text{max}_{x_1}(V),\text{max}_{x_2}(V),\text{max}_{x_3}(V)) \, .
\end{align}
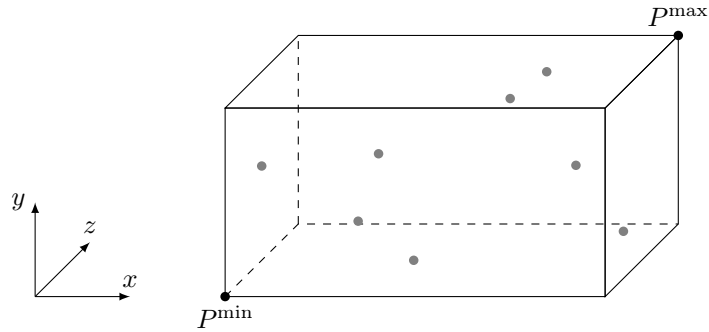
\begin{figure}[htb]
\begin{center}
\begin{tikzpicture}[scale = 2.5]
\pgfmathsetmacro{\cubex}{2}
\pgfmathsetmacro{\cubey}{1}
\pgfmathsetmacro{\cubez}{1}
\draw[black] (0,0,0) --  ++(-\cubex,0,0) -- ++(0,-\cubey,0) -- ++(\cubex,0,0) -- cycle;
\draw[black] (0,0,0) -- ++(0,0,-\cubez) -- ++(0,-\cubey,0) -- ++(0,0,\cubez) -- cycle;
\draw[black] (0,0,0) -- ++(-\cubex,0,0) -- ++(0,0,-\cubez) -- ++(\cubex,0,0) -- cycle;
\draw[black,dashed] (-\cubex,-\cubey,0) --  ++(0,0,-\cubez) -- ++(\cubex,0,0);
\draw[black,dashed] (-\cubex,-\cubey,-\cubez) --  ++(0,\cubey,0);
\draw[-latex] (-3,-1,0) -- ++(0,0,-.75) node[above] {$z$};
\draw[-latex] (-3,-1,0) -- ++(.5,0,0) node[above] {$x$};
\draw[-latex] (-3,-1,0) -- ++(0,.5,0) node[left] {$y$};
\fill (-\cubex,-\cubey,0) circle (.025) node[below] {$P^{\text{min}}$};
\fill (0,0,-\cubez) circle (.025) node[above] {$P^{\text{max}}$};
\fill[color=black!50] (-\cubex,-.5,-.5) circle (.025);
\fill[color=black!50] (-.5,0,-.5) circle (.025);
\fill[color=black!50] (-1.2,-\cubey.,-.5) circle (.025);
\fill[color=black!50] (0,-.75,-.25) circle (.025);
\fill[color=black!50] (-.25,-.4,-.25) circle (.025);
\fill[color=black!50] (-.75,-.2,-.65) circle (.025);
\fill[color=black!50] (-\cubex/2,-.\cubey/2,\cubez/2) circle (.025);
\fill[color=black!50] (-1.3,-.6,0) circle (.025);
\end{tikzpicture}
\end{center}
\caption{Axis aligned bounding box.}
\label{bittens_fig:cub}
\end{figure}Obviously, $P_{\text{min}}$ and $P_{\text{max}}$ depend on the choice of the coordinate system. Choosing the standard basis in $\mathbb{R}^3$ leads to the class of axis aligned bounding boxes  (see figure \ref{bittens_fig:cub}). Allowing for different bases $\mathcal{B}'$ an oriented bounding box
\begin{align}
B_{\mathcal{B}'}(V) = \left( P^\text{min}(\mathcal{T}_{\mathcal{B}'}V),\, P^\text{max}(\mathcal{T}_{\mathcal{B}'}V) \right)\,.
\label{bittens_seq:obb}
\end{align}
can be defined.\\
 Applying principle component analysis, shown in section \ref{bittens_sec:pca}, to the triangulation $T=(V,S)$ of the convex hull \cite{barber1996quickhull} of a body in $\mathbb{R}^3$ and using the obtained basis $\mathcal{B}^T$ as an input for the orientated bounding box 
\begin{align}
B_{\mathcal{B}^T}(V) = \left(P^ \text{min}(\mathcal{T}_{\mathcal{B}^T}V),\, P^\text{max}(\mathcal{T}_{\mathcal{B}^T}V) \right)\,,
\end{align}
the result will most likely fit the body tightly.
\subsection{Hierarchical decomposition}
%%%%%%%%%%%%%%%%%%%%%%%%%%%%%%%%%%%%%%%%%%%%%%%%
In the latter, an algorithm was defined for the creation of tight-fitting OBBs around a body in $\mathbb{R}^3$. A top-down hierarchy is used for the generation of a series of nested OBBs with gradually smaller volume. Therefore, a parent OBB $B^P_{(\star)}$ including the complete body is computed. The subdivision rule implemented then splits the bounding box with a plane orthogonal to the longest axis of the box in the center of mass. If the longest axis cannot be subdivided, the second longest axis is chosen. For the resulting two distinct regions two child OBBs $B^{C_1}_{(\star)}$ and $B^{C_2}_{(\star)}$ can be computed. This procedure may be repeated as long as a dividable convex hull of the fraction of the underlying geometry can be computed. The parent and child OBBs can be organized efficiently in a binary tree-structure. In figure \ref{bittens_fig:AABBvsOBB} two steps of this procedure for AABBs and OBBs are shown using the example of an endoprosthesis. As can been seen, the OBBs converge faster than AABBs to the shape of the underlying geometry.
%%%%%%%%%%%%%%%%%%%%%%%%%%%%%%%%%%%%%%%%%%%%%%%%
%%%%%%%%%%%%%%%%%%%%%%%%%%%%%%%%%%%%%%%%%%%%%%%%

\section{Raycasting geometric primitives in $\mathbb{R}^3$}
\label{bittens_sec:ray}
%%%%%%%%%%%%%%%%%%%%%%%%%%%%%%%%%%%%%%%%%%%%%%%%
%%%%%%%%%%%%%%%%%%%%%%%%%%%%%%%%%%%%%%%%%%%%%%%%
In computer graphics the use of ray-surface or line-surface intersection tests, needed in a variety of related problems such as volume rendering,  is termed raycasting \cite{pfister1999volumepro}. A Ray can be defined by the tuple
\begin{align}
R = (\vect{o},\vect{d})\,, \quad \vect{o},\vect{d}\in\mathbb{R}^3\,,
\end{align}
with the origin $\vect{o}$ and the direction $\vect{d}$ of the Ray.\\
In the following section the intersection of a ray with two different classes of geometric primitives is outlined: 1. the rectangular cuboid and 2. the tetrahedron.

%%%%%%%%%%%%%%%%%%%%%%%%%%%%%%%%%%%%%%%%%%%%%%%%
%\subsection{Ray with oriented bounding box}
\subsection{Intersection of ray and rectangular cuboid}
\label{bittens_sec:rayobb}
%%%%%%%%%%%%%%%%%%%%%%%%%%%%%%%%%%%%%%%%%%%%%%%%
The class of rectangular cuboids is congruent with the class of oriented bounding boxes. Therefore we restrict ourselves to the determination of the intersection of a ray with an oriented bounding box. Via the map 
\begin{align}
R^\star=\mathcal{T}_{\mathcal{B}'}R = (\mathcal{T}_{\mathcal{B}'}\vect{o},\mathcal{T}_{\mathcal{B}'}\vect{d}) 
\end{align}
the ray $R$ can easily be transformed into the coordinate system of the OBB reducing the model problem to the collision of a ray $R^\star$ with an AABB. For this problem class, the slab method, first proposed in \cite{kay1986ray} is an easy and fast solution.

\subsubsection{Slab method.} 
\begin{figure}[htb]
\begin{center}
\begin{tikzpicture}[scale = 1.4]
\scriptsize
\draw[line width = 0.8] (0,0) rectangle (3,2) node[pos=.5] {AABB};

\draw[dashed] (-2,0) node[below] {$y=P^{\text{max}}_y$} -- (0,0) ;
\draw[dashed] (3,0) -- (5,0)  ;

\draw[dashed] (-2,2) node[above] {$y=P^{\text{min}}_y$} -- (0,2);
\draw[dashed] (3,2) -- (5,2);

\draw[dashed] (0,4) -- (0,2);
\draw[dashed] (0,0) -- (0,-2) node[left] {$x=P^{\text{min}}_x$};

\draw[dashed] (3,4) -- (3,2) ;
\draw[dashed] (3,0) -- (3,-2) node[right] {$x=P^{\text{max}}_x$};

\draw[-latex,red!60!black] (2,4) node[above right] {$\vect{R}_2$} -- (5,1);
\draw[-latex,green!60!black] (-.5,4) node[above right] {$\vect{R}_1$} -- (5,-1.5);
\fill[green!60!black] (-.5,4)  circle (1pt);
\fill[red!60!black] (2,4) circle (1pt);
 
\begin{scope}[xshift=-1.5cm, yshift=-0.5cm]
\draw[-latex, line width=1.] (0,0) -- (1,0) node[below] {$x$};
\draw[-latex, line width=1.] (0,0) -- (0,-1) node[right] {$y$};
\end{scope}

\draw[fill=black] (3,3) circle (1pt) node[above right] {$t^{\text{max}}_y(\vect{R}_2)$};
\draw[fill=black] (4,2) circle (1pt) node[above right] {$t^{\text{min}}_x(\vect{R}_2)$};

\draw[fill=black] (1.5,2) circle (1pt) node[above right] {$t^{\text{min}}_x(\vect{R}_1)$};
\draw[fill=black] (3,.5) circle (1pt) node[above right] {$t^{\text{max}}_y(\vect{R}_1)$};
\draw[fill=black] (3.5,0.) circle (1pt) node[above right] {$t^{\text{max}}_x(\vect{R}_1)$};
\draw[fill=black] (0,3.5) circle (1pt) node[above right] {$t^{\text{min}}_y(\vect{R}_1)$};

\draw [decorate,decoration={brace,amplitude=8pt},xshift=-4pt,yshift=0pt]
(-2,0)--(-2,2)  node [black,midway,xshift=-0.75cm] 
{x-slab};

\draw [decorate,decoration={brace,amplitude=8pt},xshift=0pt,yshift=-4pt]
(3,-2) -- (0,-2)  node [black,midway,yshift=-0.5cm] 
{y-slab};

\fill (0,2) circle (1.5pt) node[above right] {$P^{\text{min}}$};
\fill (3,0) circle (1.5pt) node[below left] {$P^{\text{max}}$};

\end{tikzpicture}
\end{center}
\caption{Intersection-test of two Rays $\vect{R}_1$ and $\vect{R}_2$ with a 2D-AABB.}
\label{bittens_fig:slabs}
\end{figure}
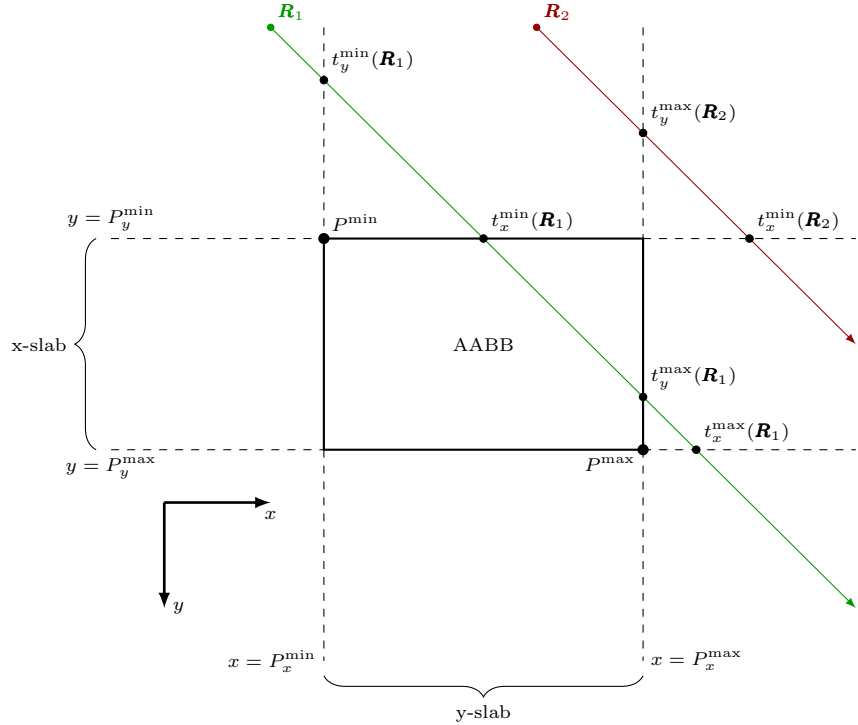

For the explanation of the slab-method the two dimensional example shown in figure \ref{bittens_fig:slabs} will be used. The procedure itself can be easily extended to three or more dimensions.\\ A ray R can also be expressed with following equation
\begin{align}
\vect{R}(t) = \vect{o} + t \cdot \vect{d}\,,\quad t\in\mathbb{R}^+\,,
\end{align}
which is a simple linear equation.
The bounding box can be represented as the intersection of two slabs, the x-slab, defined  by two parallel lines $y=P^{\text{min}}_y$ and $y=P^{\text{max}}_y$ both perpendicular to the y-axis, and the y-slab, defined  by two parallel lines $x=P^{\text{min}}_x$ and $x=P^{\text{max}}_x$ both perpendicular to the x-axis. The intersection of the ray with the x-slab can be calculated by
\begin{align}
t^{\text{min}}_x = \frac{(P^{\text{min}}_y-o_y)}{d_y} \quad \text{and} \quad t^{\text{max}}_x = \frac{(P^{\text{max}}_y-o_y)}{d_y}\,.
\end{align}
Accordingly, the intersection of the ray with the y-slab is calculated by
\begin{align}
t^{\text{min}}_y = \frac{(P^{\text{min}}_x-o_x)}{d_x} \quad \text{and} \quad t^{\text{max}}_y = \frac{(P^{\text{max}}_x-o_x)}{d_x}\,.
\end{align}
The ray then intersects with the bounding box, if and only if the greatest minimum $\text{max}(t^{\text{min}}_x,t^{\text{min}}_y)$ is smaller than the smallest maximum $\text{min}(t^{\text{max}}_x,t^{\text{max}}_y)$, which holds true for $\vect{R}_1$ and in contrast does not hold true for $\vect{R}_2$, both depicted in figure \ref{bittens_fig:slabs}. This algorithm allocates no memory and if $\frac{1}{d_i}$ is pre-computed, it is division-free, in addition.

%%%%%%%%%%%%%%%%%%%%%%%%%%%%%%%%%%%%%%%%%%%%%%%%
\subsection{Intersection of ray and tetrahedron} 
\label{bittens_sec:raysim}
%%%%%%%%%%%%%%%%%%%%%%%%%%%%%%%%%%%%%%%%%%%%%%%%
A tetrahedron can be decomposed into its set of four 2-dimensional faces. Therefore we restrict ourselves to the intersection of a ray with a 2-simplex. %Each other case, also for higher dimensions, can then be derived directly by simple combinatorics.\\
For the determination of the intersection of a ray with a 2-simplex in $\mathbb{R}^3$ the Möller–Trumbore intersection algorithm \cite{moller2005fast} can be used as follows.

\paragraph{Möller–Trumbore intersection algorithm.}\label{sec:muellertrompone}
A representation based on barycentric coordinates of a triangle $s$ with vertices $\hat{\vect{p}}$, $\hat{\vect{q}}$ and $\hat{\vect{r}}$ can be given by the set of equations:
\begin{align}
\nonumber\mathbf{s}(u,v) = (1-u-v)\cdot\hat{\vect{p}}+ u\cdot\hat{\vect{q}}+v\cdot\hat{\vect{r}}\,, \quad u,v\in\mathbb{R}\,, \\
 u \geq 0\,,  \quad   v \geq 0 \quad \text{and} \quad u+v \leq 1 \,.
\end{align}
\begin{figure}[htb]
\begin{center}
\begin{tikzpicture}[scale=1.3]
\scriptsize
%left picture
\begin{scope}[xshift=.4cm]
\draw[-latex] (0,0,0)--(.5,0,0) ;
\draw[-latex] (0,0,0)--(0,.5,0) ;
\draw[-latex] (0,0,0)--(0,0,-.65) ;
\end{scope}
%\draw[fill=white] (1,0.5,0.25) circle (1pt);
%\draw[fill=white] (2,1,0.5) circle (1pt);
%\draw[fill=white] (1.0,1.0,-0.5) circle (1pt);
% lower arrow
\draw[-latex]  (1.33333, 0.833333, 0.0833333) -- (1.8184, 0.105726, -0.401738);
%triangle
\draw[fill=white, line width = .75] (1,0.5,0.25) node[left,xshift=1] {$\hat{\vect{p}}$}-- (2,1,0.5) node[right] {$\hat{\vect{q}}$}-- (1.0,1.0,-0.5) node[above] {$\hat{\vect{r}}$}  -- cycle;
% upper arrow
\draw[] (0.848262, 1.56094, 0.568405) -- (1.33333, 0.833333, 0.0833333);
\fill[] (0.848262, 1.56094, 0.568405) circle (1pt) node[above] {$\vect{R}$};
%middle picture: translate
\begin{scope}[xshift=3.8cm]
\draw[-latex] (0,0,0)--(.5,0,0) ;
\draw[-latex] (0,0,0)--(0,.5,0) ;
\draw[-latex] (0,0,0)--(0,0,-.65) ;
% lower arrow
\draw[-latex]  ($(1.33333, 0.833333, 0.0833333)-(1,0.5,0.25)$) -- ($(1.8184, 0.105726, -0.401738)-(1,0.5,0.25)$);
%triangle
\draw[fill=white, line width = .75] (0,0,0) -- ($(2,1,0.5)-(1,0.5,0.25)$) node[right] {$\hat{\vect{q}}-\hat{\vect{p}}$}-- ($(1.0,1.0,-0.5)-(1,0.5,0.25)$) node[right, xshift = 0, yshift = 3] {$\hat{\vect{r}}-\hat{\vect{p}}$}  -- cycle;
% upper arrow
\draw[] ($(0.848262, 1.56094, 0.568405)-(1,0.5,0.25)$) -- ($(1.33333, 0.833333, 0.0833333)-(1,0.5,0.25)$);
\fill[] ($(0.848262, 1.56094, 0.568405)-(1,0.5,0.25)$) circle (1pt) node[above] {$\vect{R}-\hat{\vect{p}}$};
\end{scope}
\begin{scope}[xshift=7cm]
\draw[-latex] (0,0,0)--(1.5,0,0) node[above] {$u$};
\draw[-latex] (0,0,0)--(0,1.,0) node[left] {$t$};
\draw[-latex] (0,0,0)--(0,0,-1.5) node[above] {$v$};
\draw[-latex] (.333,0,-.333) -- (.333,-.75,-.333);
\draw[fill=white, line width = .75] (0,.0,0) -- (1,0,0) node [below] {$1$} -- (0,0,-1) node [above left] {$1$} -- cycle;
\draw[] (.333,1,-.333) -- (.333,0,-.333);
\fill (.333,1,-.333) circle (1pt) node[above right] {$ \mathbf{M}^{-1} \left[\vect{R}- \hat{\vect{p}} \right] $};
\end{scope}
\path[-latex] (2.3,0.4) edge[bend left] node[above] {translate} (3.3,0.4);
\path[-latex] (5.6,0.4) edge[bend left] node[above] {$\mathbf{M}^{-1}$} (6.6,0.4);
%\draw[-latex, shorten >=1ex, shorten <=1ex] (2.2,0) edge[bend right] (3.2,0);
\end{tikzpicture}
\end{center}
\caption{Translation and change of basis of the ray and the triangle \cite{moller2005fast}.}
\label{bittens_fig:trint}
\end{figure}
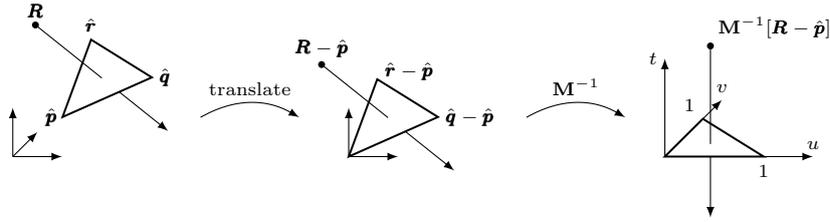
The intersection of a ray $\vect{R}(t)$ with a triangle $\mathbf{s}(u,v)$ can be found by solving the equation
\begin{align}
\nonumber\vect{R}(t) &= \mathbf{s}(u,v) \\
\Rightarrow \vect{o}+t \cdot \vect{d} &= (1-u-v)\cdot\hat{\vect{p}}+ u\cdot\hat{\vect{q}}+v\hat{\vect{r}}\,.
\end{align}
Rearranging the terms 
\begin{align}
-t\cdot {\vect{d}} + u \cdot (\hat{\vect{q}}-\hat{\vect{p}}) + v\cdot (\hat{\vect{r}}-\hat{\vect{p}})= {\vect{o}} - \hat{\vect{p}}\,,
\end{align}
which can be seen as the translation of $\hat{\vect{p}}$ to the origin (see figure \ref{bittens_fig:trint}), yields the linear system of equations
\begin{align}
\underbrace{\begin{bmatrix} -\vect{d}, & \hat{\vect{q}}-\hat{\vect{p}}, & \hat{\vect{r}}-\hat{\vect{p}} \end{bmatrix}}_{\mathbf{M}}\begin{bmatrix}
t\\u\\v
\end{bmatrix} = \vect{o} - \hat{\vect{p}}\,,
\label{bittens_eq:M}
\end{align}
where $\mathbf{M}^{-1}$ can be seen as the transformation of $\vect{R}$ and $\mathbf{s}$, such that $\mathbf{s}$ is the unit-triangle shown in figure \ref{bittens_fig:trint}.\\
Denoting $\vect{E}_1 = \hat{\vect{q}}-\hat{\vect{p}}$, $\vect{E}_2 = \hat{\vect{r}}-\hat{\vect{p}}$ and $\vect{T}=\vect{o}-\hat{\vect{p}}$ the solution to (\ref{bittens_eq:M}) can be obtained using Cramer's rule
\begin{align}
\begin{bmatrix}
t\\u\\v
\end{bmatrix} = \frac{1}{(\vect{d}\times \vect{E}_2)\cdot \vect{E}_1} \begin{bmatrix}
(\vect{T}\times \vect{E}_1)\cdot \vect{E}_2\\(\vect{D}\times \vect{E}_2)\cdot \vect{T}\\(\vect{T}\times \vect{E}_1)\cdot \vect{d}
\end{bmatrix}\,.
\end{align}
The ray $\vect{R}(t)$ then intersects with the triangle $\mathbf{s}(u,v)$ at $\vect{o}+t \cdot \vect{d}$, if and only if $u>0$, $v>0$, $t>0$ and $u+v\leq 1$.
%\subsection{Ray with $3$-simplex}
%For the intersection of a ray with a  3-simplex, the algorithm explained in section \ref{bittens_sec:raysim} can be used to calculate the intersection of the ray with the boundary of the 3-simplex, obtaining at most two intersection points $t_{\text{in}}$ and $t_{\text{out}}$.  

\section[Positional relationship]{Positional relationship of a discrete point with a finite element}
Finding the points of intersection for a ray and a finite element with higher polynomial order is not trivial  and there are only a few sources available that provide a solution. Among them, \cite{wiley2004ray} and \cite{uffinger2010interactive} describe raycasting algorithms for high-quality visualization of finite elements. Since in this work we are not interested in visualizing finite elements the problem is reduced to the detection of the positional relationship of a discrete point $\vect{x}$ and a finite element, as mentioned earlier. With this simplification the problem reduces to the search of the local coordinates $\bs{\xi}(\vect{x})$ dependent on the point $\vect{x}$, given in the global coordinate system.
\subsection{Global-to-local iteration}
For a given finite element with nodal coordinates $\hat{\vect{x}}^i$, $i=1,...,n_\text{nodes}$, associated shape functions $N_i(\bs{\xi})$  and a point $\vect{x}$, given in global coordinates, the element-local coordinates $\boldsymbol{\xi}$ need to be found, such that
\begin{align}
\vect{f}(\boldsymbol{\xi})=\mathbf{N}_i(\boldsymbol{\xi}) \, \hat{\mathbf{x}}^i - \mathbf{x} = \boldsymbol{0}\,.
\end{align}
This problem can be solved by the Newton-Raphson algorithm 
\begin{align}
\boldsymbol{\xi}^{(n+1)} = \boldsymbol{\xi}^{(n)} - \mathbf{J}(\boldsymbol{\xi}^{(n)})^{-1} \, f(\boldsymbol{\xi}^{(n)})\,,
\end{align}
with  $\tens{J}$ being the Jacobian matrix with its components
\eq{
\tcomp{J}_{ij} = \frac{\partial f_i}{\partial \xi_j} (\boldsymbol{\xi})\,,  \quad {i = 1,\ldots,3\,,\,j=1,\ldots,3}\,,
} an appropriate initial guess $\boldsymbol{\xi}_0$ and a convergence criterion
\eq{
\frac{||\boldsymbol{\xi}^{(n+1)}-\boldsymbol{\xi}^{(n)}||}{||\boldsymbol{\xi}^{(1)}||} < \varepsilon_\text{tol}\,,
}
with a user-defined tolerance $\varepsilon_{\text{tol}}$.
\subsection{In-hull test}
Presuming $\bs{\xi}(\vect{x})$ is known, the quantity of interest $q$ can be interpolated in terms of shape functions from a finite element post-processing result as $q(\vect{x}) = N_i(\bs{\xi}(\vect{x}))\cdot \hat{q}^i$. What remains is to check whether the local coordinates lie within the bounds of the reference element. Since all commonly used types of finite elements are usually convex, a simple test whether $\bs{\xi}(\vect{x})$ lies within the convex hull of the reference finite element is sufficient. For the class of tetrahedral elements as used in this thesis an in-hull test, taking advantage of the barycentric coordinate system, can be performed by
\eq{
\sum\limits_i \xi_i \leq 1 \quad \text{and} \quad \xi_i\geq 0\,, \quad i=1,\ldots,3\,,
}
for example.
\section{Attenuation law}
The initial intensity $I_{\text{in}}$ of a beam of e.g. electromagnetic radiation decreases as it passes through a volume of matter. The \textit{Beer-Lambert} law \cite{swinehart1962beer} relates this attenuation as
\begin{align}
I_{\text{out}} = I_{\text{in}} \, e^{-\int\mu(\vect{x}) \, \text{d}x}\,,
\end{align}
where $I_{\text{out}}$ is the resultant intensity and $\mu$ is the \textit{linear attenuation coefficient}. As an example, the linear attenuation coefficients for compact and cancellous bone are stated as $2.251$ $\text{cm}^{-1}$ and $0.716$ $\text{cm}^{-1}$, respectively,  in \cite{schneider1985getrennte}.
\section{X-ray generation}
For the X-ray setup, a rather simply model is chosen here with an orthographic projection and monochromatic X-ray beams. In an orthographic projection the X-ray source is a plane with equally distributed parallel X-rays while monochromatic means each X-ray beam shares the same energy. For more information about possible X-ray simulation setups, the reader is referred to \cite{sujar2017gvirtualxray}, for example.\\
A summary of the X-ray simulation process can then be stated as follows:
\begin{enumerate}
\item An axis aligned bounding box enclosing the complete finite element model is generated, whose surfaces can serve as emission plates. 
\item A tree $G$ of hierarchical-structured oriented boundary boxes is generated to subdivide the finite element model.
\item For every X-ray $\vect{R}_i$ sent through the model, the following steps are performed:
\begin{enumerate}
\item Find all leaves of $G$, which intersect with $\vect{R}_i$.
\item Introduce numerical integration scheme $\bar{\vect{R}}_{ij} = \vect{o}+t_j\cdot\vect{d}$ by sampling the ray equidistantly with a desired resolution.
\item \label{en:ray}For each discrete point $\bar{\vect{R}}_{ij}$, perform the global-to-local iteration only on the finite elements inside the leaf bounding boxes.
\item Perform numerical integration on the quantity of interest, e.g. \eq{ I_{\text{out}} = I_{\text{in}} \, \text{exp}\left[{-\sum\limits_jw_j\mu(\varrho(\vect{R}_{ij}))}\right]\,.}
\end{enumerate}
\end{enumerate}
Albeit is possible to introduce hierarchical levels in the tree $G$ until each leaf is filled only with one finite element, it is not always sensible to do so. It was found to be more efficient to create leaves with less than 10 elements and then order them for each ray $\vect{R}_i$ by a linear guess performed by the Möller-Trumbore algorithm, introduced in \ref{sec:muellertrompone}.
\section{Numerical examples}
All methods described in this chapter were implemented within a self-developed framework, written in the {\julia} language \cite{bezanson2017julia}. In order to demonstrate the functionality of the method and the implementation, two numerical examples are derived here. The first one is a coarsely discretized ball with constant density and the second one is a finer discretized disk with a quadratic density distribution assigned to it.
\subsection{Ball with constant density}
\label{sec:virtxrayexamplesphere}
\begin{figure}[htb]
\begin{center}
\includegraphics[width=.36\textwidth]{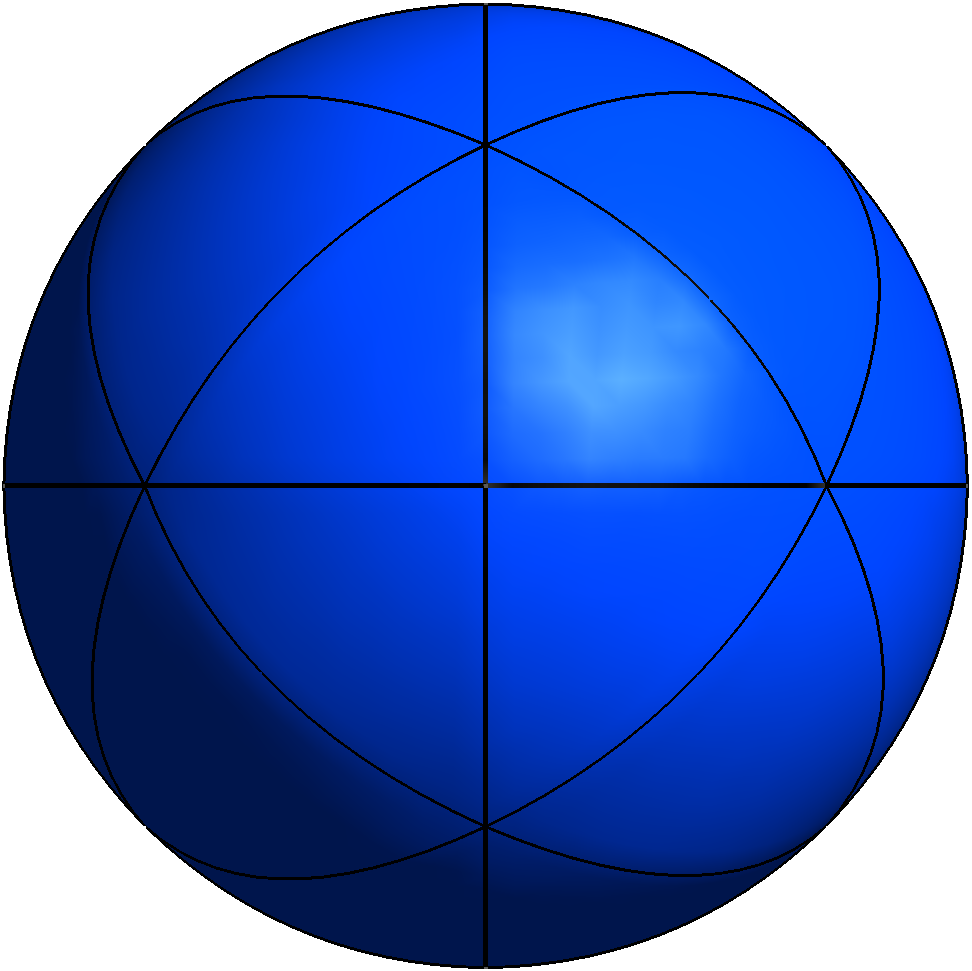}
\hspace{.1\textwidth}
\includegraphics[width=.35\textwidth]{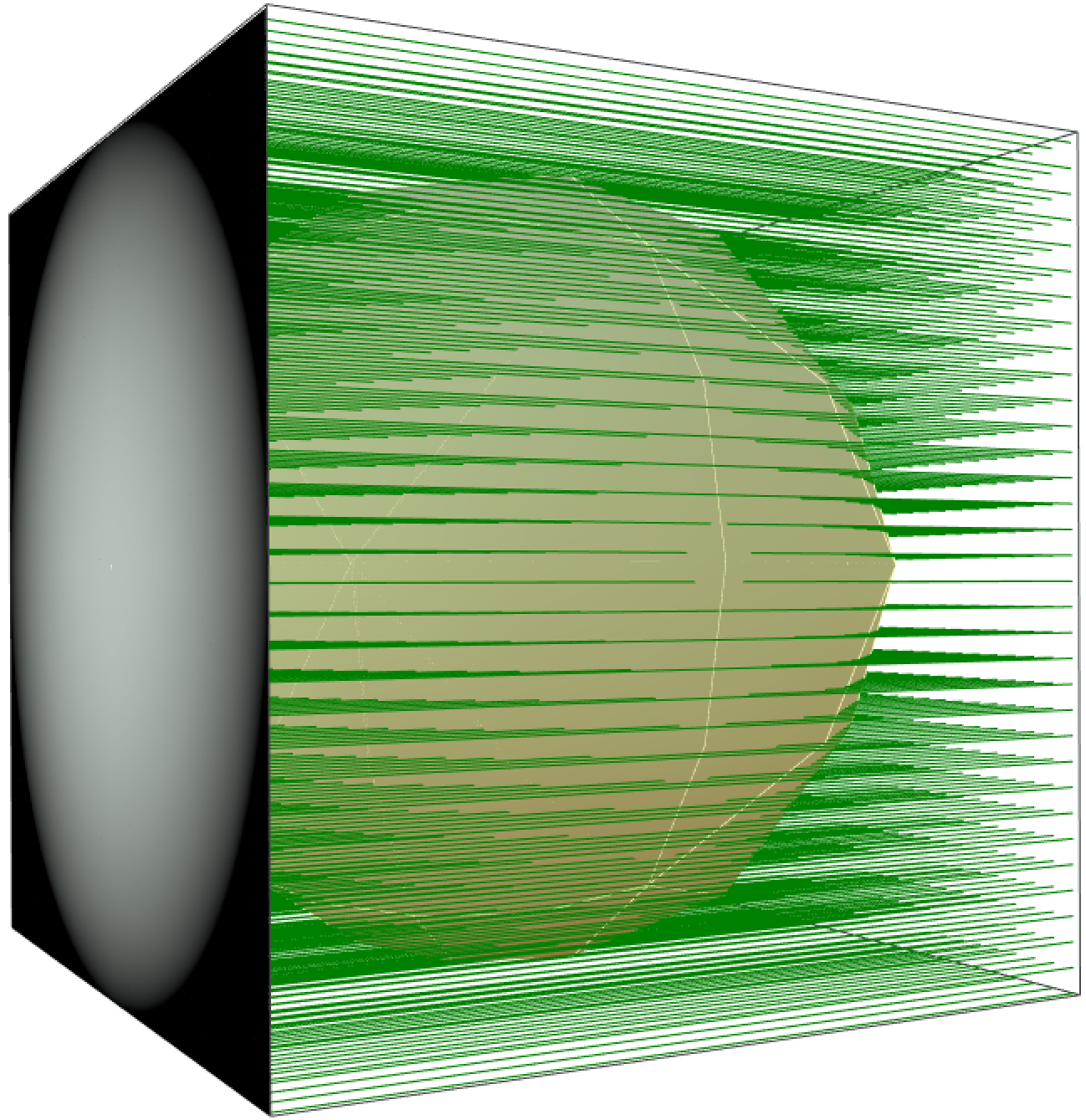}
\caption{Ball discretized by 50 quadratic tetrahedral elements (left picture) and the model setup with ball, initial bounding box, X-ray projection plane and selected rays assembled in one scene (right picture).}
\label{fig:sphere_model_setting}
\end{center}
\end{figure}

As a first benchmark problem, a zero-centered ball $S:x^2+y^2+z^2=r^2$, $x,y,z,r\in \mathbb{R}$ of radius $r=1$~cm is discretized coarsely with 50 quadratic 10-node tetrahedral elements (see figure \ref{fig:sphere_model_setting}).
A constant density of $\varrho = 1 \frac{\text{g}}{\text{cm}^3}$ is assigned to each element. An orthographic X-ray projection-plane is set up, as can be seen in figure \ref{fig:sphere_model_setting}, using the X-ray algorithm as an integration for the density of the discretized version of the ball. 23.668 rays per $\text{cm}^2$ are used to sample the discretized ball, which results in 94.864 rays in total, where the density is sampled along each ray with the same frequency, resulting in 29.218.112 sample points in total.
\begin{figure}[htb]
\centering
\begin{tikzpicture}[every node/.style={inner sep=2,outer sep=0}]
\node (label) at (-2.25,-1.85)[]{
        \includegraphics[width=.35\textwidth]{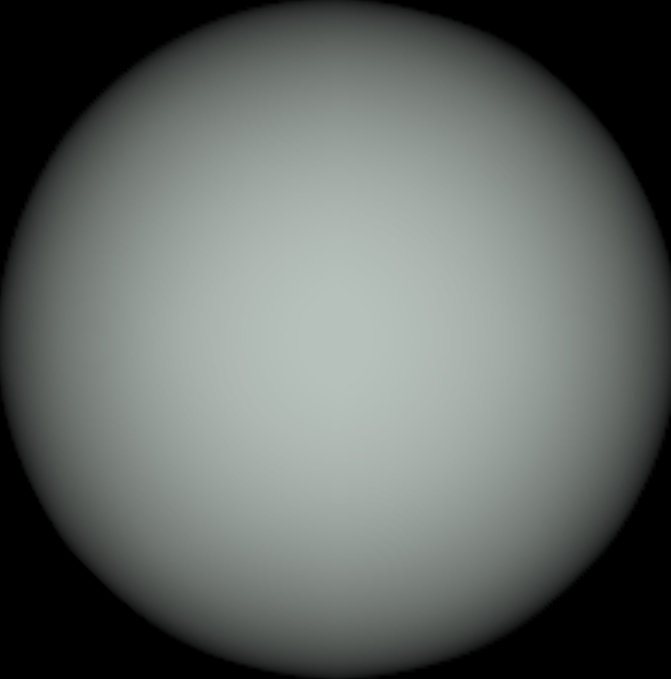}
      };\hspace{.25cm}
\pgfplotscolorbardrawstandalone[ 
    colormap/blackwhite,
    colorbar,% horizontal,
    point meta min=0,
    point meta max=2,
    colorbar style={
        height=3.9cm
        }
        ]
\end{tikzpicture}
\hspace{.01\textwidth}
\begin{tikzpicture}[every node/.style={inner sep=2,outer sep=0}]
\node (label) at (-2.25,-1.85)[]{
        \includegraphics[width=.35\textwidth]{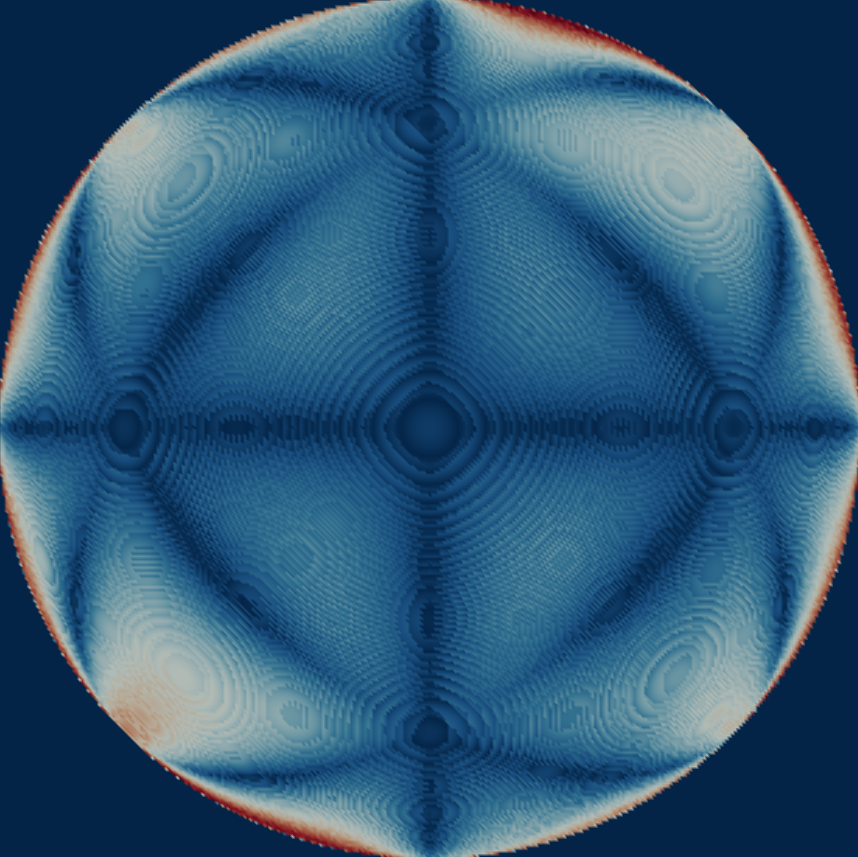}
      };\hspace{.25cm}
\pgfplotscolorbardrawstandalone[ 
    colormap/jet,
    colorbar,% horizontal,
    point meta min=0,
    point meta max=0.08,
    colorbar style={
	%yticklabel={\footnotesize \pgfmathparse{10^\tick}\pgfmathprintnumber\pgfmathresult},
    	%scaled ticks=false,
        %width=10cm,
        height=3.9cm,
        }
        ]
\end{tikzpicture}
\caption{X-ray projection of the ball with density displayed in $\nicefrac{\text{g}}{\text{cm}^2}$ (left picture) and logscale plot of absolute error in $\nicefrac{\text{g}}{\text{cm}^2}$ (right picture).}
\label{fig:sphere_error}
\end{figure}
In figure \ref{fig:sphere_error} the resulting X-ray image as well as the absolute error can be seen. The X-ray image looks as expected with a maximum density of 1.995 $\nicefrac{\text{g}}{\text{cm}^2}$ at the middle of the picture, as the X-ray passes the full diameter here. Integration once again over the projected density results in a mass of $4.135\,\text{g}$, which has an error of approximately $1.3\,\%$ error when compared to the analytically calculated mass $m=\varrho\, \frac{4}{3} \pi r^3 \approx 4.189\,\text{g}$. The absolut error, which can be seen as the sum of the discretization error of the ball and the discretization error of the rays, is measured at the location of each ray $R_i$ by comparing the projected density with the integral  $\int_{r_{i,\text{in}}}^{r_{i,\text{out}}} \varrho\, \text{d}x$, where $r_{i,\text{in}}$ and $r_{i,\text{out}}$ are the intersection points of the ray with the analytical ball $S$. It can be seen that the error is the smallest\begin{figure}[htb]
\begin{center}
\includegraphics[width=0.325\textwidth]{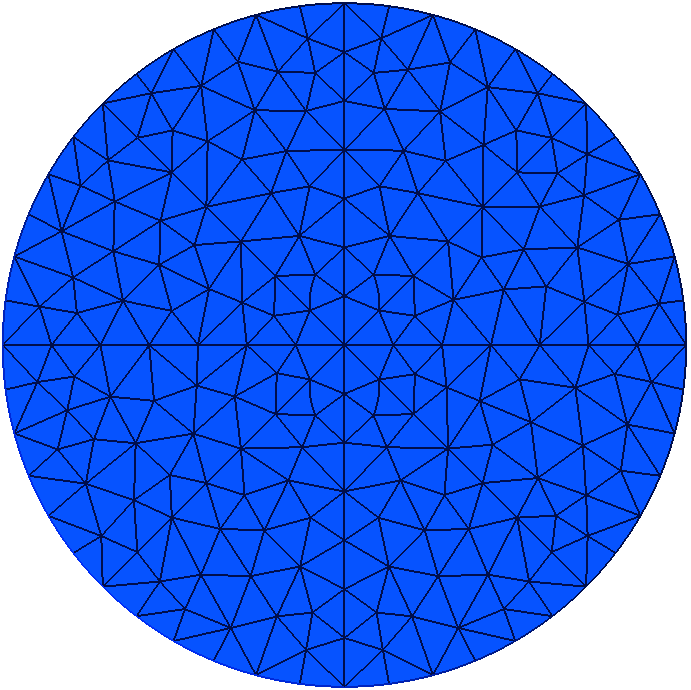}
\hspace{.01\textwidth}
\begin{tikzpicture}
\begin{axis}[width=.375\textwidth,every axis plot post/.append style={
  mark=none,domain=-0:1,samples=100,smooth},
    % All plots: from -2:2, 50 samples, smooth, no marks
  axis x line=bottom, % no box around the plot, only x and y axis
  axis y line=left, % the * suppresses the arrow tips
   xlabel     = {r $ \left[ \text{cm} \right] $},
   ylabel     = {density $ \left[ \frac{\text{g}}{\text{cm}^3} \right] $},
  enlargelimits=upper] % extend the axes a bit to the right and top
  \addplot {-4*(x-0.5)^2+2.0};
\end{axis}
\end{tikzpicture}
\caption{Cylinder discretized by 2143 quadratic tetrahedral elements (left picture) and density distribution depending on the radius $r$ (right picture).}
\label{fig:cylinder}
\end{center}
\end{figure}
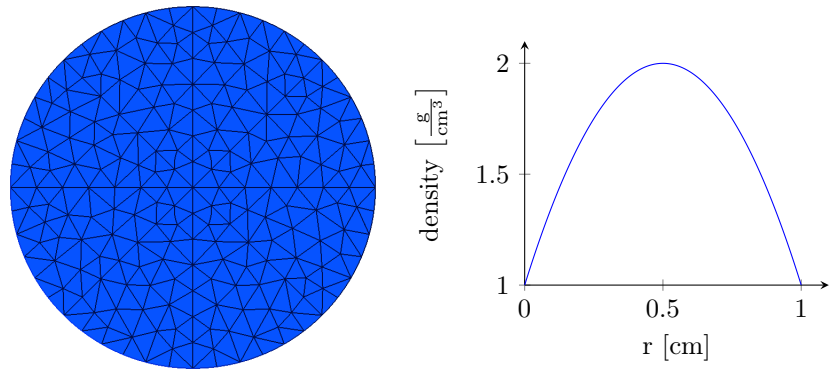 where the vertices of the 10-node tetrahedral elements are clipped to the surface of the ball, followed by the edges and finally the faces. As expected the largest error occurs at the edge of the projected circle. In addition concentric circles can be seen all over the picture, which can be interpreted as a sign of the depth discretization of the rays.
\subsection{Cylinder with quadratic density distribution}
As second test, a cylinder with radius $r=1\,\text{cm}$ and height $h=0.1\,\text{cm}$ is discretized with 2143 quadratic 10-node tetrahedral elements and a quadratic density distribution along the radius $\varrho(r) = -4(r-0.5)^2+2$ is assigned to the model (both shown in figure \ref{fig:cylinder}).\begin{figure}[htb]
\centering
\begin{tikzpicture}[every node/.style={inner sep=2,outer sep=0}]
\node (label) at (-2.25,-1.85)[]{
        \includegraphics[width=.35\textwidth]{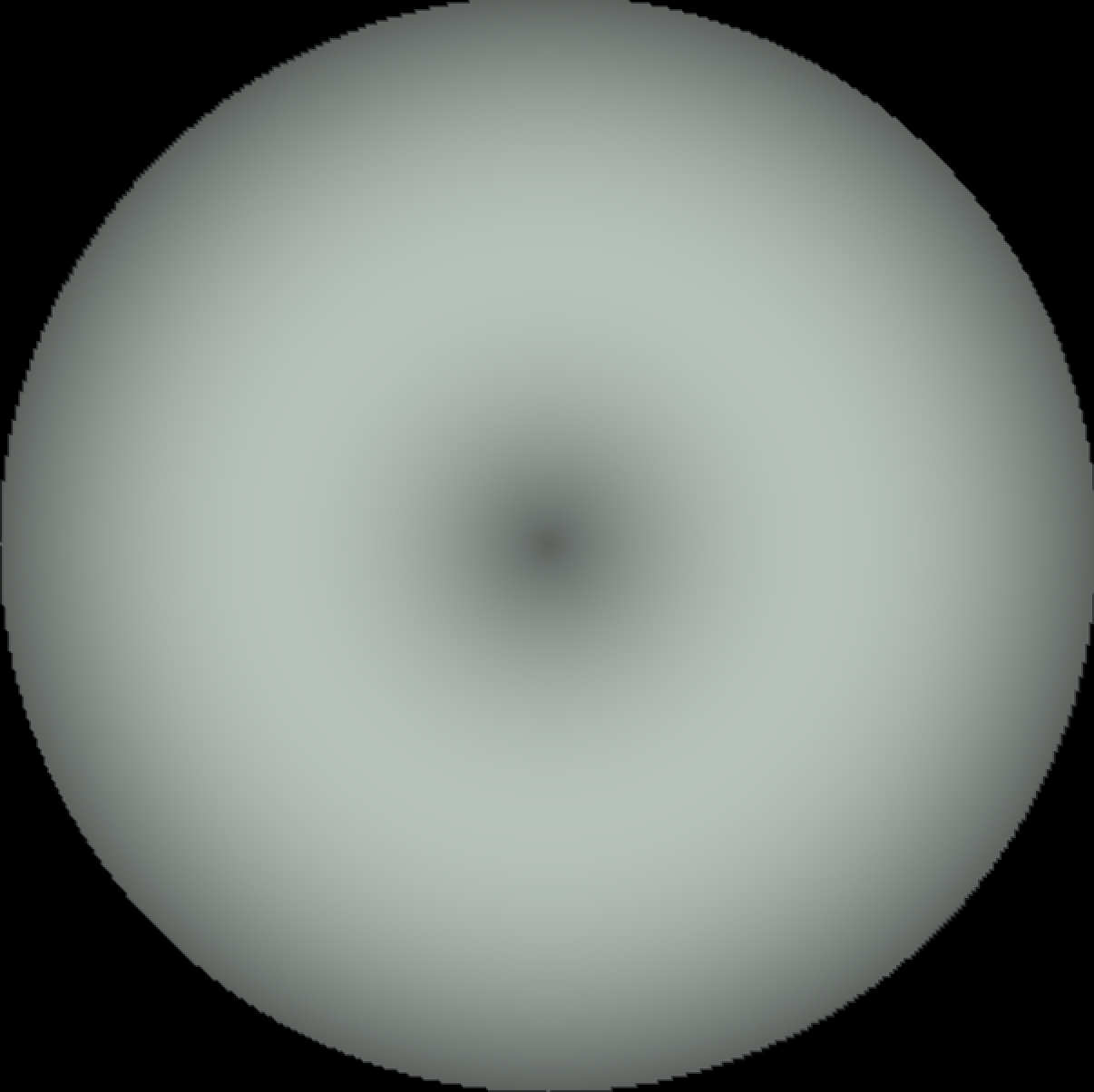}
      };\hspace{0.25cm}
\pgfplotscolorbardrawstandalone[ 
    colormap/blackwhite,
    colorbar,% horizontal,
    point meta min=0,
    point meta max=0.2,
    colorbar style={
  	ytick={0,0.1,0.2},
        height=3.9cm
        }
        ]
\end{tikzpicture}
\hspace{.01\textwidth}
\begin{tikzpicture}[every node/.style={inner sep=2,outer sep=0}]
\node (label) at (-2.25,-1.85)[]{
        \includegraphics[width=.35\textwidth]{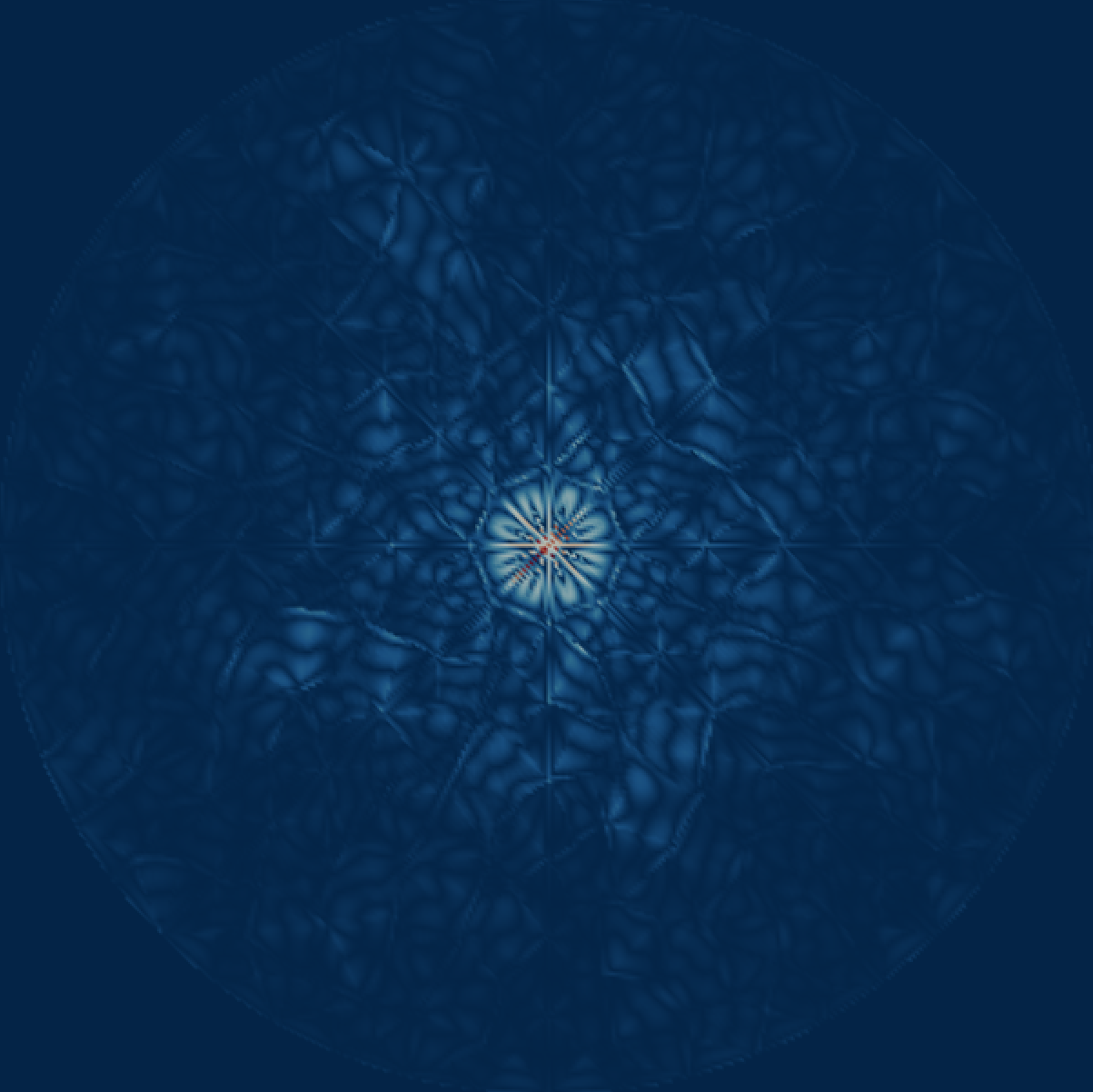}
      };\hspace{0.25cm}
\pgfplotscolorbardrawstandalone[ 
    colormap/jet,
    colorbar,% horizontal,
    point meta min=0,
    point meta max=0.00015,
    colorbar style={
%	yticklabel={\footnotesize \pgfmathparse{10^\tick}\pgfmathprintnumber\pgfmathresult},
    	%scaled ticks=false,
        %width=10cm,
        height=3.9cm,
        }
        ]
\end{tikzpicture}
\caption{X-ray projection of the cylinder with density displayed in $\nicefrac{\text{g}}{\text{cm}^2}$ (left picture) and plot of absolut error in $\nicefrac{\text{g}}{\text{cm}^2}$ (right picture).}
\label{fig:cylinder_error}
\end{figure} The X-ray setup is analogues to the latter example with the exception that there are less samples in the depth direction as the geometry and the density exhibit no variation in that direction and therefore the integration is exact even with just one sample along each ray.\\
In figure \ref{fig:cylinder_error} the projected density and the absolute error can be seen. The projected density looks as expected, which once again demonstrates the function of the X-ray simulation. The absolute error is at least two magnitudes lower as in the first example. This can be explained by the simplified geometry, which, as already mentioned, exhibits exact integration per ray and by the finer mesh which renders the discretization of the cylinder almost exactly.

	\section{Bibliography}
	\bibliography{biblio}
	\bibliographystyle{unsrtdin1}

\end{document}